\documentclass[a4paper]{jpconf}
\usepackage{graphicx}
\begin{document}
\title{The stifness of the supranuclear equation of state (once again)}

\author{J. E. Horvath and R. A. de Souza}

\address{Departamento de Astronomia, IAG-USP, S\~ao Paulo University, 05579-090 S\~ao Paulo SP, Brazil}

\ead{foton@iag.usp.br}

\begin{abstract}
We revisit the present status of the stiffness of the supranuclear equations of state, particularly the solutions that 
increase the stiffness in the presence of hyperons, the putative transition to a quark matter phase 
and the robustness of massive compact star observations.
\end{abstract}

\section{The advent of high masses}
Astrophysics is not a precisely a fairy tale, but if it was, one could read a future imaginary book on compact stars starting as:

\bigskip
``Once upon a time, everybody believed that a neutron star was composed of neutrons and its typical mass 
was not very different from $1.4 M_{\odot}$...'' 
\bigskip

This belief actually lasted until the first years of the 21st century, it can be found im many books and review 
articles and was part of the ``conventional wisdom'' in the high-energy field. However, as early as 1970's 
the possibility of an exotic composition (quarks) was considered. In its mildest version, quarks should 
be the relevant degrees of freedom at $\sim$ several times the nuclear saturation density and constitute 
an inner central core. In its extreme version, a form of cold quark matter (strange quark matter) was hypothesized 
to be the true ground state of strong interactions and grow to compose most of the ``neutron'' stars [1-3]
Many other phases were postulated (i.e. pion and kaon condensates, etc.) and the task of both theoreticians 
and observational astrophysics was to disentangle the composition through a combination of calculations and 
observations of actual compact stars, a task which is still in the beginning (see [4,5] for reviews). 

The last few years yield a definite advance in the determination of neutron stars masses (although radii are 
still controversial and elusive [5]). As we shall discuss, there is now conclusive evidence for heavy ($\leq 2 M_{\odot}$) 
neutron stars, and therefore the softest equations of state became unrealistic. However, it is not yet clear from a 
theoretical point of view where do we stand, and what does it really mean for hyperon and quark models, two 
of the best studied alternatives for the internal composition. We shall review here some of these possibilities 
without the pretension to be extensive, but rather to exemplify the type of behavior of the equation of state 
that can comply with the observations, and what type of physical elements are required. We conclude with a 
brief critical apprisal of this problem as it stands in 2016.

\section{Observations and masses}
It is well-known that high masses (substantially above $1.4 M_{\odot}$) can be produced by stiff equations of state only, 
therefore the state-of-the-art pre-2000 or so did not discriminate among the several possibilities on this basis only.
This is why in the last years firm evidence for masses well above the former ``canonical'' value was received by the community 
as an actual novelty, possibly allowing a better comprehension of the equation of state itself in the long run.

The first important evidence for high masses was presented by Demorest et al [6] with the univocal detection of 
the Shapiro delay in the system containing the source PSR J1614-2230. The precision of the fit allowed a robust (and clean)
determination of a value $1.97 \pm 0.04 \, M_{\odot}$ well accepted today. A more conventional approach (optical observations 
of the companion, allowing a detrmination of radial velocities + modeling of the companion WD) allowed Antoniadis et al. [7] 
to calculate a mass of $2.01 \pm 0.04 \, M_{\odot}$ for PSR J0348+0432, reinforcing this high mass findings. 
Very recent works have also contributed to this problem, for instance, Falanga et al. [8] reported a mass of 
$2.12 \pm 0.16 \, M_{\odot}$ for the HMXB Vela X-1 and $1.96 \pm 0.19 \, M_{\odot}$ for 4U 1700-37. The quest for 
the highest mass continues because of its enormous implications, presently the record is held by PSR J1748-2021B in NGC 6440 [9]
with a whopping $2.74 \pm 0.21 \, M_{\odot}$, although this extreme value needs confirmation. 

The whole issue of the mass distribution has been a subject of interest recently. The inclusion of high mass objects and 
application of Bayesian tools prompted Valentim, Rangel and Horvath [10] to claim at least a bimodal distribution 
with peaks at $1.37  M_{\odot}$ and $1.73  M_{\odot}$. Somewhat different values, but nevertheless firm evidence for more 
than one peak has been presented by Zhang et al [11], \"Ozel et al [12] and Kiziltan, Kottas and Thorsett [13].
Using the latest data Ozel et al. updated their values to $1.39  M_{\odot}$ and $1.8  M_{\odot}$, and also inferred a 
possible maximum mass of $2.15 M_{\odot}$. The important point here is that a bin with massive NS has been identified, 
even though its exact shape is still uncertain. Recent reviews on this issue can be found in [5] and [14].
While still far from the Rhoades-Ruffini limit [15] there is no question that masses above $2 M_{\odot}$ should be 
accomodated by theoretical proposals, and this new limit brings considerable discussion for the stifness of the 
equation of state, as we discuss below.

\section{Hyperons? Which hyperons?}

The inclusion of hyperons in dense matter is an anathem lasting at least 40 years. Hyperons are known to exist (!) 
and a natural part of the baryonic sector. Within a non-relativistic approach and even allowing a considerable 
uncertainty about their actual coupling to neutrons and protons, hyperons are believed to appear at around 
$2-3 \rho_{0}$, and soften considerably the equation of state (see M. Baldo's talk in these Proceedings). Thus, 
hyperons are considered ``nasty beasts'' in nuclear physics at least regarding this issue. Many equations 
of state present in the literature did not include hyperons at all or neglect their interaction with neutrons and 
protons and/or treat the matter in the mean-flield approach, which is known to produce stiffer equations of 
state almost by construction.

With the confirmation of high-mass neutron stars the issue of the presence of the hyperons gained a new twist, and 
became known as the ``hyperon puzzle''. In a few words this name refers to the quandary posed by the (inevitable?) 
presence of hyperons which would preclude precisely the high masses. 

The so-called hyperon puzzle has some sources which can be readily identified as a source of problems. 
The first is to check what is really included and what is not when a definite scheme of calculations is given, including 
all the interaction terms and three-body forces (see Pederiva's talk in these Proceedings). In particular, it is 
important to check whether three-body forces can be ``tuned'' to give more repulsion ($C_{T} < 1$ in nuclear 
physics standard notation). The second issue is the old problem of computed vs. measured hypernuclei binding 
energies. While calculations are not particularly difficult,  experiments are scarce and probe a limited range 
only. This lack of reliable input is a source of considerable uncertainty regarding the influence of interactions on 
the hyperon appearance in dense matter, noting also that for the intersting regime the constraints are even 
waeker than for the nuclear matter case (see I. Bombaci review for a full appraisal of this issue [16]).

\subsection{A model calculation as an example}

As a definite example treating the hyperon fields we shall discuss our own results employing the $\sigma - \omega - \rho - \delta - \phi$ 
presented in Gomes et al. [17]. The model includes the full baryonic octet and meson interactions that bear its name. 
The presence of derivative couplings somehow ``sums up'' the many body forces discussed above.  The most relevant part of the 
Lagrangian reads

\begin{equation}
{\sum_{b}}{\overline \psi_{b}}{\biggl( 1 + {{g_{\sigma b} \sigma + g_{\sigma^{*} b} {\sigma^{*}} + {1\over{2}} g_{\delta b} {\bf {\tau}. {\delta}}} \over{\zeta m_{b}}} \biggr)}^{-\zeta} m_{b} \psi_{b} 
\end{equation}

The effective masses become

\begin{equation}
m_{\lambda b}^{*} = {\biggl( 1 + {{g_{\sigma b} \sigma + g_{\sigma^{*} b} {\sigma^{*}} + {1\over{2}} g_{\delta b} {\bf {\tau}. {\delta}}} \over{\lambda m_{b}}} \biggr)}^{-\lambda}
\end{equation}

with $\lambda = \xi, \kappa, \eta, \zeta$ and $b$ denotes the components of the baryonic octet.. This model attempts to describe many-body forces by parametrizing them in terms of 
non-linear interactions. It also uses the concept of `´naturalness'' for the copupling constants. 
Thus, its results go beyond the usual treatment of H-n interactions and may display a solution to the hyperon puzzle.

The equation of state derived by Gomes et al. [17] and employed here is shown in Fig. 1. Here $\zeta$ has been identified as the most important parameter of all 
for the behavior (slope) of the equation of state, once the properties of nuclear matter at the saturation point have been employed to fix the remaining set. 
Decreasing $\zeta$ produces a stiffening of the equation of state for a range of acceptable values of the former.

\begin{figure}[h]
\includegraphics[width=18pc]{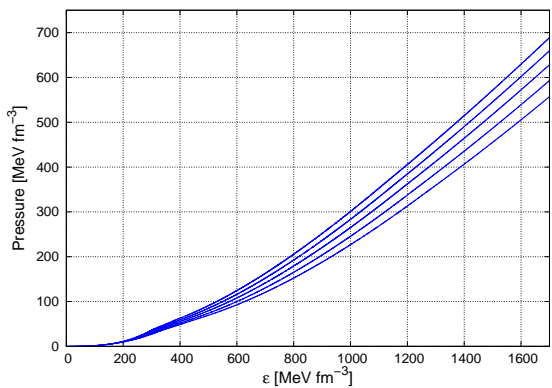}\hspace{2pc}%
\begin{minipage}[b]{14pc}\caption{\label{label}The equation of state of the $\sigma - \omega - \rho - \delta - \phi$ model. The curves have been generated by fixing all parameters 
but $\zeta$ and varying the latter. From bottom to top the adopted values were 8.5, 7.1, 5.9, 4.9 and 4 $\times 10^{-2}$.}
\end{minipage}
\end{figure}

With the aim of investigating the appearance of quark matter in the central core, we have constructed equilibrium phase transitions employing the stiff 
equation of state for cold quark matter (three flavors in chemical equilibrium) 
denominated MFTQCD due to Franzon et al. [18]. The latter begins by performing a mean-field approximation 
and separating soft and hard gluos according to Foga\c ca and Navarra [19]. A dynamical mass of the gluon $m_{G}$ is generated by the mean field 
expectation value of the condensates of order 2, while those of the oder 4 produce a constant term identified as a contribution to the vacuum energy.
The value of the latter and the quotient $\xi = g/m_{G}$ are the parameters of the model. 
The important point here is that the equation of state is ``bag-like'' and stiff enough to have some chance to be present at very high pressures.

The equation of state is shown in Fig. 2. For the more stiff versions we have attempted to construct a phase trasition {\it \`a la Maxwell} ignoring the 
conservation of many charges, as suggested by the work of Yasutake et al. [20]  who showed that it is almost impossible to produce 
a large coexistence region within realistic models of the interface.

\begin{figure}[h]
\includegraphics[width=18pc]{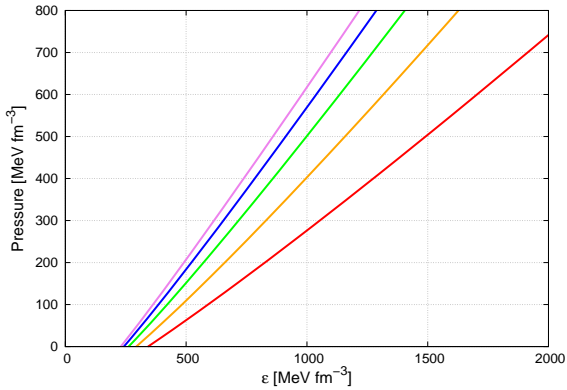}\hspace{2pc}%
\begin{minipage}[b]{14pc}\caption{\label{label}The equation of state of the MFTQCD cold quark matter phase. The vacuum term has been fixed as $B_{MFTQCD}=90 MeV/fm^{3}$,  
which is consistent with known hadronic physics and the quotient $\xi$ varied from bottom to top to be, 1.5, 3, 4.5, 6 and 7.5 $\times 10^{-4}$. Note that the stifness of the equation 
of state {\it decreases} with $\xi$.}
\end{minipage}
\end{figure}

This attempt is shown in Figs. 3 and 4. As seen, the existence of a quark phase is not easy, precisely because the $\sigma - \omega - \rho - \delta - \phi$
produces enough repulsion to postpone the appearance of quark matter, at least for most values of the parameter $\zeta$.

\begin{figure}[h]
\includegraphics[width=18pc]{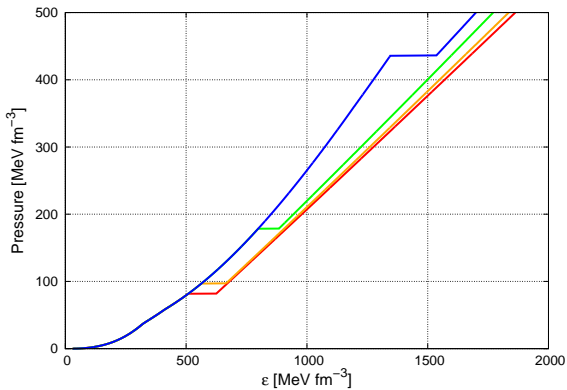}\hspace{2pc}%
\begin{minipage}[b]{14pc}\caption{\label{label}Maxwell construction for the transition between the $\sigma - \omega - \rho - \delta - \phi$ and the MFTQCD quark phase. 
For definitness the $B_{MFTQCD}=90 MeV/fm^{3}$ and $\zeta = 5.9 \times 10^{-2}$ and $\xi$ has been varied from 4 $\times 10^{-4}$ (lowest coexistence pressure) to 
9 $\times 10^{-4}$ (highest coexistence pressure). The stiffness of the quark phase controlled by the latter parameter and determining the appearance or absence of 
the phase is apparent. }
\end{minipage}
\end{figure}

\begin{figure}[h]
\includegraphics[width=18pc]{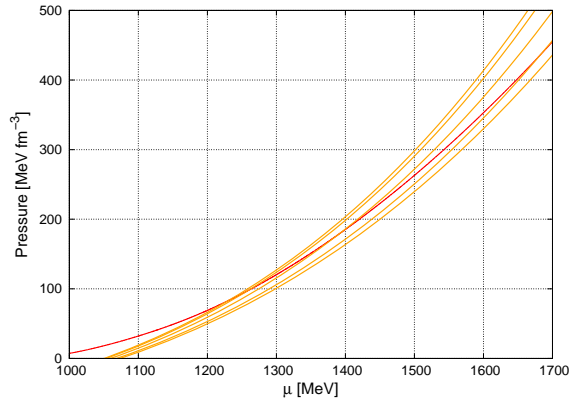}\hspace{2pc}%
\begin{minipage}[b]{14pc}\caption{\label{label}An example of equilibrium between $\sigma - \omega - \rho - \delta - \phi$ and the quark phase described by the MFTQCD for a favorable 
set of parameters in which the curves cross.}
\end{minipage}
\end{figure}

The equations of state just described were employed to integrate the TOV equations and generate theoretical non-rotating stellar sequences. As a 
result (Fig. 5), we verified that the maximum mass $M_{max}$ is barely above the last determined value $\sim 2 M_{\odot}$ for a limited set of the 
parameters. In other words, this class of hyperonic equations of state is in some trouble to reproduce the highest masses, and even more, if they happen 
to be correct, it is almost impossible that a quark core can develope inside them. While one may think of constructing an even stiffer EoS for the 
quark phase, it is diffcult to imagine that vacuum+interaction terms in cold quark matter can change this feature.

\begin{figure}[h]
\includegraphics[width=18pc]{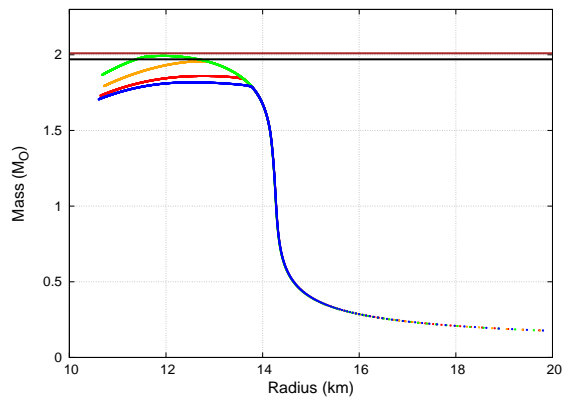}\hspace{2pc}%
\begin{minipage}[b]{14pc}\caption{\label{label}The stellar sequences emerging from the integration of the equations of state. The constraint $M_{max} > 2 M_{\odot}$ is barely satisfied }
\end{minipage}
\end{figure}

\section{Conclusions}

We revisited the issue of the stiffness of the equation of state within a limited set of possibilities prompted by the study of the hyperonic sector.
We do not yet know how large an actual compact star mass could be, just that $\sim 2 M_{\odot}$ objects have been confirmed. Higher masses 
would be even more difficult to accomodate (of course), and the physics of the ultradense matter could be approaching the Rhoades-Ruffini limit 
dangerously, without a clear physical reason for that.

Since we know that without the inclusion of hyperons the maximum masses of the sequence can go above $\sim 2 M_{\odot}$ quite easily, 
one strange solution to the hyperon puzzle could be the ``hyperonless'' hyperonic matter, in which hyperons are supressed by some convenient 
mechanism. A conventional alternative to this would be to have enough repulsion from them. However, and stated quite boldly but pointing to the 
trend which is clearly present in our calculations, the latter 
possibility would relegate the study of cold quark matter to the status of an academic problem (unless that SQM composes all compact 
objects and we have to admit from scratch that we never experienced the true ground state of hadronic interactions in the lab...).

\bigskip
\noindent
{\bf Acknowledgements}
We would like to acknoledge the Local Organizing Committee for the organization of a nice CSQCD Meeting. JEH has been 
supported by the Fapesp Agency (Process 2013/26258-4) and CNPq, Brazil (Process 305328/2013-1).

\section*{References}

\noindent
[1] Bodmer, A 1971 {\it Phys. Rev. D} {bf 4} 1601

\noindent
[2] H. Terazawa, Tokyo U. Report. (1979) INS–336

\noindent
[3] Witten, E 1984 {\it Phys. Rev. D}{\bf 30} 272

\noindent
[4] Lattimer, J 2012 {\it Annu. Rev. Nucl. Part. Sci.} {\bf 62} 485 

\noindent
[5] \"Ozel, F, Freire, P 2016 {\it Annu. Rev. Astron. Astrophys.}{\bf 54} 401

\noindent
[6] Demorest, PB, Pennucci, T, Ransom, S M, Roberts, MSE, Hessels, JWT 2010 {\it Nature}{\bf 467} 1081

\noindent
[7] Antoniadis, J, Freire PCC, Wex N, Tauris TM, Lynch RS, van Kerkwijk MH, Kramer M, Bassa C, 
Dhillon VS, Driebe T, Hessels JWT, Kaspi VM, Kondratiev VI, Langer N, Marsh TR, McLaughlin MA, 
Pennucci TT, Ransom SM, Stairs IH, van Leeuwen J, Verbiest JPW, Whelan DG 2013 {\it Science 340} 448

\noindent
[8] Falanga, M, Bozzo, E, Lutovinov, A, Bonnet-Bidaud, JM, Fetisova, Y, Puls, J 2015 {\it A\&A}{\bf 577} A130

\noindent
[9] Freire PCC, Ransom SM, Bégin S, Stairs IH, Hessels JWT, Frey LH, Camilo F 2008 {\it ApJ}{\bf 675} 670

\noindent
[10] Valentim R, Rangel E, Horvath JE 2011 {\it MNRAS}{\bf 414} 1427

\noindent
[11] Zhang, CM, Wang J, Zhao YH, Yin HX, Song LM, Menezes DP, Wickramasinghe DT, Ferrario L, Chardonnet P 2011 {\it A\&A}{\bf 527} A83 

\noindent
[12] \"Ozel F, Psaltis D, Narayan R, Santos Villareal A 2012 {\it ApJ}{\bf 757} A55

\noindent
[13] Kiziltan B, Kottas A, De Yoreo M, Thorsett SE 2013 {\it ApJ}{\bf 778} A66

\noindent
[14] Horvath, JE, Valentim, R 2016 arXiv:1607.06981

\noindent
[15] Rhoades, CE, Ruffini, R 1974 {\it Phys. Rev. Lett.}{\bf 32} 324

\noindent
[16] Bombaci, I 2016 arXiv:1601.05339

\noindent
[17] Gomes, RO,  Dexheimer, V, Schramm, S, Vasconcellos, CAZ 2015 {\it ApJ}{\bf 808} 8

\noindent
[18]  Franzon, B, Foga\c ca, DA, Navarra, FS, Horvath, JE 2012 {\it Phys. Rev. D}{\bf 86} 065031

\noindent
[19] Foga\c ca, DA, Navarra, FS 2011 {\it Phys. Lett. B}{\bf 700} 236

\noindent
[20] Yasutake, N, Chen, H, Maruyama, T, Tatsumi, T 2016 {\it Journal of Physics: Conference Series} {\bf 665} 012068 

\end{document}